# Scandium decorated $C_{24}$ fullerene as high capacity reversible hydrogen storage material: Insights from density functional theory simulations

*Vikram Mahamiya[a], Alok Shukla[a\*], Brahmananda Chakraborty[b,c\*]*

[a]Indian Institute of Technology Bombay, Mumbai 400076, India

[b]High pressure and Synchrotron Radiation Physics Division, Bhabha Atomic Research Centre, Bombay, Mumbai, India-40085

[c]Homi Bhabha National Institute, Mumbai, India-400094

email: shukla@phy.iitb.ac.in ; brahma@barc.gov.in

## ABSTRACT

Using first principles density functional theory simulations, we have observed that the scandium decorated $C_{24}$ fullerene can adsorb up to six hydrogen molecules with an average adsorption energy of -0.35 eV per $H_2$ and average desorption temperature of 451 K. The gravimetric wt % of hydrogen for the scandium decorated $C_{24}$ fullerene system is 13.02%, which is sufficiently higher than the Department of Energy, United States demand. Electronic structure, orbital interactions, and charge transfer mechanisms are explained using the density of states, spatial charge density difference plots, and Bader charge analysis. A total amount of 1.44e charge transfer from the 3d and 4s orbitals of scandium to the 2p carbon orbitals of $C_{24}$ fullerene. Hydrogen molecules are attached to scandium decorated $C_{24}$ fullerene by Kubas type of interactions. Diffusion energy barrier calculations predict that the existence of a sufficient energy barrier will prevent metal-metal clustering. *Ab-initio* molecular dynamics (A.I.M.D.) simulations confirm the solidity of structure at the highest desorption temperature. Therefore, we believe that the scandium decorated $C_{24}$ fullerene system is a thermodynamically stable, promising reversible high-capacity hydrogen storage device.

## 1. INTRODUCTION

Due to the continuous depletion of limited fuel energy sources, the sharp increase in the global population, and environmental pollution, the scientific community is looking towards alternative energy sources[1]. Hydrogen is considered one of the prominent alternative of fossil fuel sources because it has the highest energy per unit weight, high natural abundance, and environmentally friendly behavior[2–4]. There are many limitations in hydrogen storage for practical fuel application[3,5]. Large, heavy pressure tanks are needed for hydrogen storage in gaseous form, and transportation of these bulky tanks is not an easy task[2]. Safety is also one of the major concerns. The liquid form of hydrogen is not affordable due to the extra liquefaction cost[4]. The storage of hydrogen in solid form is ideal, provided that hydrogen molecules have suitable adsorption energy, desorption temperature, and gravimetric weight percentage (wt %). The adsorption energy of hydrogen should be in between -0.2 eV to -0.7 eV, and at least 6.5 gravimetric wt % of hydrogen should be stored in hydrogen storing media as per the guidelines of department of energy, United States(DOE-US)[6,7].

People have studied different materials such as metal hydrides[8–10], metal alloys[11–14], metal-organic frameworks[15,16], zeolites[17] for hydrogen storage. Instability of the structure at high temperature, high desorption temperature, metal-metal cluster formation, and lower wt % of hydrogen are some of the difficulties with these materials that should be taken care of for an ideal hydrogen storage substrate. Carbon nanomaterials are promising hydrogen storage substrates because of their larger surface area and relatively lower molecular mass than most of the other materials. People have studied different kinds of carbon nanomaterials such as graphene, graphyne (2d), carbon nanotubes (1d), and fullerene (0d) for hydrogen storage applications[18–27]. Carbon nanomaterials in their pure form are not potentially applicable for hydrogen storage as the interaction is only due to the weak van der Waals (vdW) forces between hydrogen and carbon nanomaterials at room temperature[21,24,25,28–30]. Transition metal decorated carbon nanomaterials adsorb hydrogen near room temperature and do not dislodge the hydrogen at small elevated temperatures due to the thermal fluctuations[19,31]. Hydrogen storage properties of yttrium decorated single-wall carbon nanotube (SWCNT) are studied by Chakraborty et al.[32]. They observed that each yttrium atom can adsorb 6 $H_2$ molecules, and the system is thermodynamically stable at a very high temperature 700 K. Modak et al.[33] have reported a comparative performance for the hydrogen storage by Y, Zr, Nb, and Mo decorated single-wall carbon nanotube. They have proposed that the transition metals with the minimum number of d electrons in the outer shell are better hydrogen adsorbers. Hydrogen storage capability reduces as the number of d electrons in the outer shell increases. It was also observed that Y and Zr doped SWCNT is metallic while Nb and Mo doped SWCNT is semiconducting. Hydrogen storage in Ti-doped single-wall carbon nanotube is studied by Yildirim et al.[34]. They have proposed that Ti-doped SWCNT can adsorb 4 $H_2$ molecules leading to 8 wt% of hydrogen. Hydrogen storage in Ca doped graphene is studied by Ataca et al.[35]. They have predicted 8.4 wt% of hydrogen. Lebon et al.[36] have reported hydrogen storage in Ti-doped graphene nanoribbons using dispersion corrected DFT. Zr doped graphene system for hydrogen storage is studied by Yadav. et al.[37], and a high wt% of 11 was reported. Recently Chakraborty et al.[38] have studied hydrogen storage in Ti-doped psi graphene system. They have found high wt % of hydrogen 13.14%, for this advanced member of graphene family. Yttrium doped graphyne structure with 10% hydrogen uptake is studied by Gangan et al.[39]. Zhang et al.[40] have observed 5.8 wt % of hydrogen in Y doped $B_{40}$, but the desorption temperature of hydrogen for their system was found to be lesser than room temperature. Yildirim et al.[41] have studied hydrogen storage in scandium and titanium doped $C_{60}$ fullerene and predicted up to 7% of hydrogen uptake for their system. Zhao et al.[42] have found that scandium interacts with $C_{60}$ and $C_{48}B_{12}$ fullerene and produces a stable organometallic fullerene structure. They have achieved up to 9% of hydrogen uptake for their system. Hydrogen storage in Pd and Co-decorated $C_{24}$ fullerene was reported by Soltani et al.[43], but they have not investigated the solidity of the structure at high temperatures. Sathe et al.[44] have investigated hydrogen storage in Ti-doped $C_{24}$ fullerene recently. They observed that a single titanium atom adsorbs 4 $H_2$ molecules with 10.5 wt % of hydrogen for their system, but the issue was the higher maximum desorption temperature (562 K).

Hydrogen storage capacity of metal decorated carbon nanostructures is also explored experimentally and it was observed that metal decorated carbon nanostructures are potential hydrogen storage candidates. Hydrogen storage capacity of palladium nanoparticles decorated multi-walled carbon nanotubes (MWCNTs) was studied by Mehrabi et al.[45] using laser ablation and chemical reduction methods. They have reported gravimetric uptake of 6 wt% of hydrogen for their system. Hydrogen storage capacity of Ni and Al-doped graphene composites is studied by Gu et al.[46]. They have reported maximum hydrogen storage of 5.7% for their system. Tarasov et al.[47] have predicted 6.5 wt% of reversible hydrogen storage capacity for magnesium composites with nickel and graphene-like material.

The $C_{24}$ fullerene structure has $O_h$ point group symmetry. It was experimentally detected in 2001 by laser ablation on a graphite surface[48]. Hydrogen storage in scandium decorated $C_{24}$ fullerene is not studied so far. Scandium is the lightest transition metal element, and it has sufficient empty d-orbitals to bind the hydrogen molecules by Kubas interaction[49–51]. Some of the previous studies on scandium doped $C_{60}$ fullerene[41,42] structure also suggest that scandium is one of the most suitable transition metal dopants for the hydrogen storage application on fullerene molecules.

We have explored the hydrogen adsorption and desorption phenomenon in scandium decorated $C_{24}$ fullerene with $O_h$ point group symmetry in this paper. We have calculated that the average binding energy of hydrogen molecules is -0.35 eV. The value of average desorption temperature is 451 K, which lies in the suitable range for the fuel cell application. We report a very high gravimetric wt % 13.02 for the scandium decorated $C_{24}$ fullerene structure. We have calculated the diffusion energy barrier for the scandium atom, which is sufficiently higher than the thermal energy of scandium at the highest desorption temperature. The solidity of structure at high temperature is checked by the *ab-initio* molecular dynamics (A.I.M.D.) simulations, which predict that the chances for the metal atom to dislodge from the fullerene structure even at the highest desorption temperature are very less. We present the density of states (DOS), partial density of states (PDOS), and charge density difference plot to understand the qualitative charge transfer mechanism. The exact amount of charge transfer is calculated by Bader charge analysis[52]. Scandium decorated $C_{24}$ fullerene structure has suitable adsorption energy, and desorption temperature for the hydrogen molecules, diffusion energy barrier for scandium atom is much higher than the thermal energy at desorption temperature, the structure is stable at desorption temperature, and hydrogen uptake is very high. Therefore, we believe that scandium decorated $C_{24}$ fullerene is a potential hydrogen storage device.

## 2. COMPUTATIONAL DETAILS

All the simulations have been performed using Density Functional Theory (DFT). Vienna Ab Initio Simulation Package (VASP)[53–56] is used for the simulations employing generalized gradient approximation (GGA) exchange-correlation functional along with the Grimme's dispersion corrections DFT-D3[57,58]. Since local density approximation (LDA) exchange-correlation functional overestimate the bonding strength of the system[59], we have used GGA exchange-correlation functional for the calculations. One unit cell of $C_{24}$ fullerene is taken in a large cubic box of size 30 Å to avoid periodic interactions. The cut-off energy of the plane waves is taken to be 500 eV. We have set a convergence limit of 0.01 eV/ Å and $10^{-5}$ eV for the force and the energy, respectively. To check the integrity of the scandium decorated $C_{24}$ fullerene structure at high temperature, we have performed the *ab-initio* molecular dynamics simulations[60]. The system was kept in a microcanonical (NVE) ensemble for 5 ps time duration in the time step of 1 fs, and the temperature is raised up to 500 K, then in a canonical ensemble for the next 3 ps at 500 K.

## 3. RESULTS AND DISCUSSIONS

### 3.1 Interaction of scandium on $C_{24}$ fullerene

We have considered one isolated unit cell of $C_{24}$ fullerene molecule for the density functional theory simulations. The relaxed structure of $C_{24}$ fullerene is presented in **Fig. 1(a).** There are two different types of C-C bonds in the relaxed structure of $C_{24}$ fullerene. One bond is in between the common face of a hexagon and a tetragon with bond length $l_{6,4} = 1.49$ Å, while the other bond is in between the common face of two hexagons having bond length $l_{6,6} = 1.38$ Å. The bond lengths of the relaxed structure are in very good agreement with the previously reported values $l_{6,6} = 1.38$ Å and $l_{6,4} = 1.50$ Å[61].

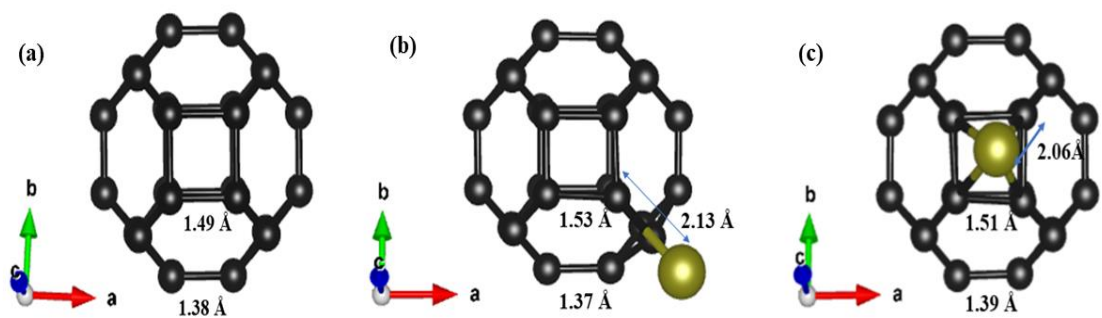

**Fig. 1** Optimized structures of (a) $C_{24}$ fullerene (b) Sc-decorated $C_{24}$ fullerene where Sc-atom is placed in front of the interfacial side of two hexagons (c) Sc-decorated $C_{24}$ fullerene where Sc-atom is placed on the top of the tetragon of $C_{24}$ fullerene. Black and golden colors correspond to C-atom and Sc-atoms, respectively.

Scandium atom is kept at different locations of $C_{24}$ fullerene molecule, and we have found two stable structures of scandium decorated $C_{24}$ fullerene as displayed in **Fig. 1(b)** and **Fig. 1(c).** In **Fig. 1(b),** scandium atom is attached in front of the common face of two hexagons, 2.13 Å distance away from the $C_{24}$ fullerene molecule with binging energy -3.44 eV. The scandium atom is attached in front of the tetragon of $C_{24}$ fullerene at 2.06 Å distance away from the $C_{24}$ fullerene molecule in **Fig. 1(c).** The binding energy of the scandium atom corresponding to **Fig. 1(c)** is -3.12 eV. Negative binding energy denotes the exothermic reaction which is responsible for the stability of structures. Since the binding of the scandium atom when it is attached in front of the common face of two hexagons of $C_{24}$ fullerene is stronger, we have considered this structure for hydrogen storage. Since The binding energy of the Sc atom (3.44 eV) is comparable with the experimental cohesive energy of the Sc atom (3.90 eV)[62], we have performed the *ab-initio* molecular dynamics simulations and diffusion energy barrier calculations in section. 3.6, to check whether the Sc atom remains intact on the $C_{24}$ fullerene at high desorption temperatures. We have calculated the binding energy of scandium on $C_{24}$ fullerene structure using the following formula:

$$(B.E.)_{Sc} = E_{C24+Sc} - (E_{C24} + E_{Sc}) \qquad (1)$$

Where $E_{C24+Sc}$, $E_{C24}$, and $E_{Sc}$ are the energy of Sc-decorated $C_{24}$ fullerene, energy of $C_{24}$ fullerene, and energy of isolated Sc-atom, respectively.

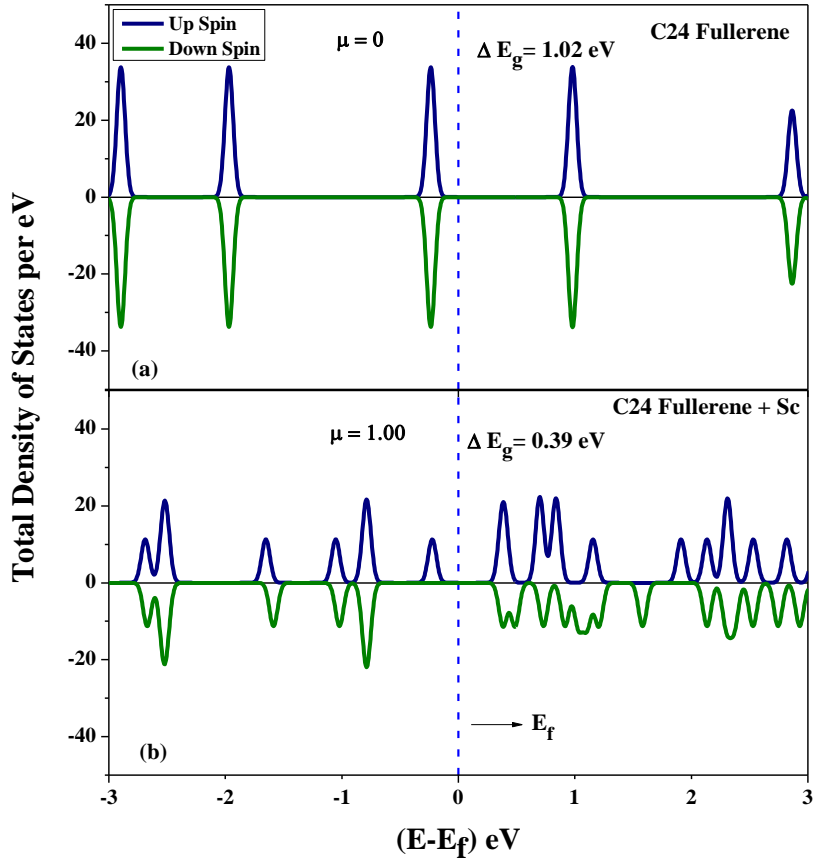

**Fig. 2** Total density of states of (a) $C_{24}$ fullerene (b) Sc-decorated $C_{24}$ fullerene. μ represent the induced magnetic moment. Fermi level is set at 0 eV.

In **Fig. 2,** we present the total density of states of $C_{24}$ fullerene and scandium decorated $C_{24}$ fullerene structures. The symmetry in the density of states of $C_{24}$ fullerene with respect to up and down spin panels is due to the non-magnetic behavior of $C_{24}$ fullerene. We can notice in **Fig. 2**, that when the scandium atom is decorated on $C_{24}$ fullerene, the density of states is not symmetric anymore, which means now the system has become magnetic. The induced magnetic character in the system is due to the unpaired d electrons of the scandium atom. The energy band gap (HOMO-LUMO gap) of $C_{24}$ fullerene structure using GGA exchange-correlation is 1.02 eV, lesser than the previously reported value 1.79 eV by Xu et al.[63]. To get the more accurate electronic density of states for $C_{24}$ fullerene, we have used hybrid functional for calculating the band gap of the $C_{24}$ fullerene structure. The band gap of the $C_{24}$ fullerene using HSE06 functional[64] was found to be 1.77 eV, which is in agreement with the literature. The total density of states of $C_{24}$ fullerene with GGA and HSE06 exchange-correlation functional are presented in **Fig.3.**

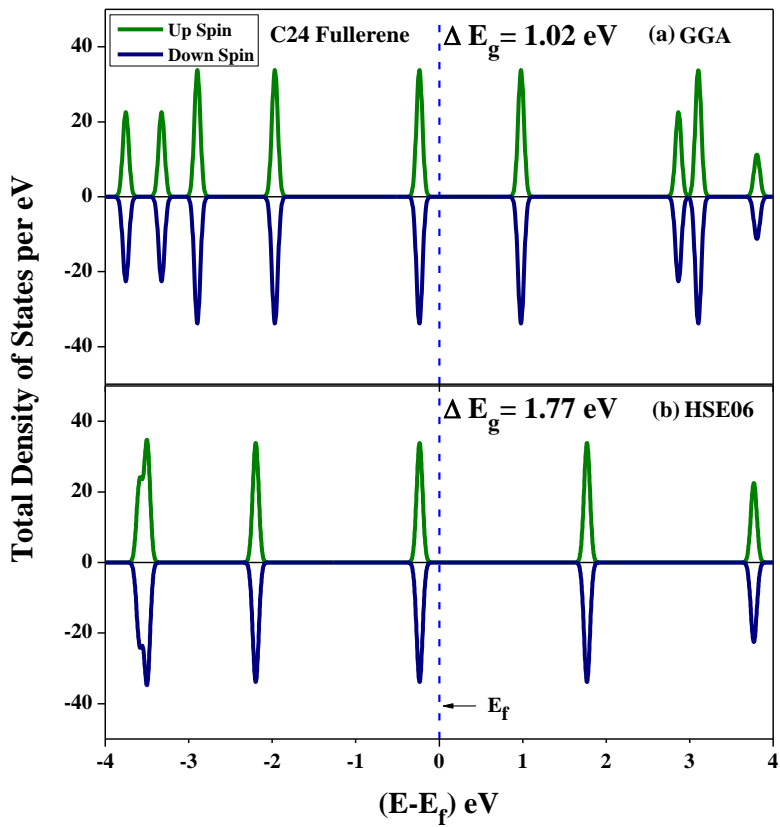

**Fig. 3 Total density of states of $C_{24}$ fullerene using (a) DFT-GGA method (b) DFT-HSE06 method. Energy band gap of $C_{24}$ fullerene increases when HSE06 exchange correlation is employed. Fermi level is set at 0 eV.**

Scandium decorated $C_{24}$ fullerene structure has an energy band gap of 0.39 eV and induced magnetic moment of 1 μB. Hence the metallicity of the $C_{24}$ fullerene increases after the decoration of the scandium atom.

### 3.2 Interactions and Bonding mechanism between scandium atom and $C_{24}$ fullerene

To explain the orbital interactions and the bonding mechanism in the scandium decorated $C_{24}$ fullerene structure, we have plotted and analyzed the partial density of states and charge density difference plots.

**Partial density of states (PDOS) analysis**

To understand the electronic structure and orbital interactions in scandium decorated $C_{24}$ fullerene, we have presented the partial density of states of C-2p orbitals of $C_{24}$ fullerene and $C_{24}$ + Sc system as presented in **Fig. 4(a)** and **Fig. 4(b),** respectively. We can notice some enhancement in the states near Fermi level in **Fig. 4(b)** compared to **Fig. 4(a),** which implies that some charge has been gained by C-2p orbitals of $C_{24}$ fullerene, when the fullerene is attached with the scandium atom. To get clearer insights regarding enhancement of electronic states of C 2p orbitals, we have plotted the partial density of states of Sc-3d orbitals for isolated scandium atom and $C_{24}$ + Sc system in **Fig. 4(c),** and **Fig. 4(d),** respectively. We can notice that there are some intense states near the Fermi level of the isolated Sc atom, which are missing when scandium is attached to $C_{24}$ fullerene. This indicates the charge loss phenomenon; hence we can conclude that some charge has been transferred from the Sc-3d orbitals to C-2p orbitals of $C_{24}$ fullerene when the scandium atom is attached to $C_{24}$ fullerene.

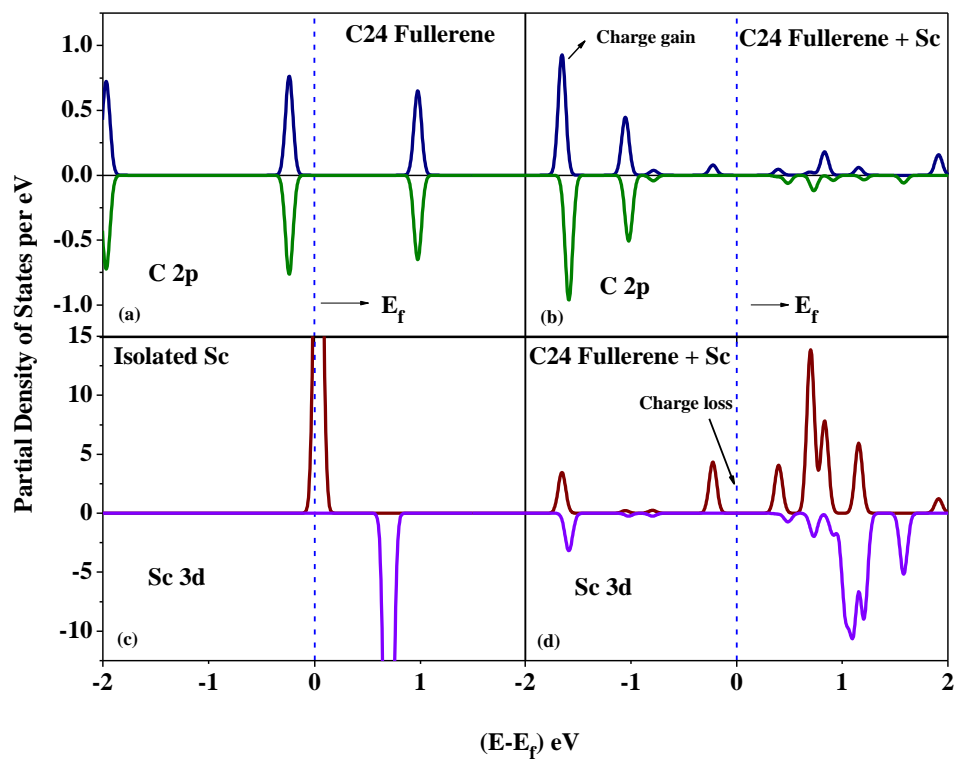

**Fig. 4** Partial density of states for (a) C-2p orbitals of $C_{24}$ fullerene. (b) C-2p orbitals of $C_{24}$ + Sc. (c) Sc-3d orbitals of isolated Sc atom. (d) Sc-3d orbitals of $C_{24}$ + Sc. Fermi level is set at zero energy value.

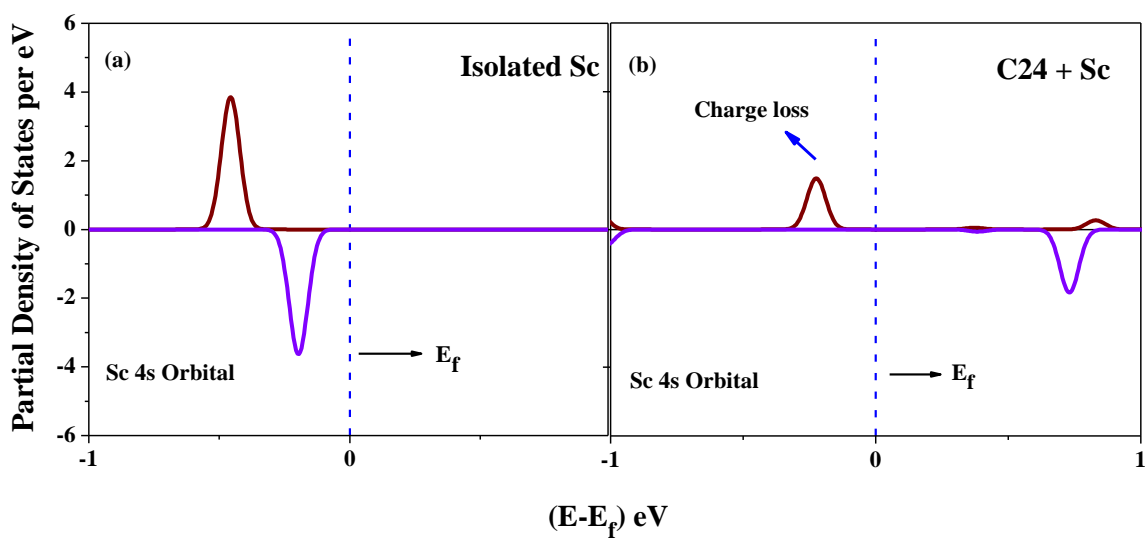

**Fig. 5** Partial density of states for (a) Sc-4s orbital of isolated Sc atom. (b) Sc-4s orbital of $C_{24}$ + Sc. Fermi energy is set at zero energy value.

We have also analyzed the partial density of states of 4s orbital of scandium atom corresponding to isolated scandium atom and $C_{24}$ + Sc system as displayed in **Fig. 5(a)** and **Fig. 5(b),** respectively. Here, we can see the reduction in the states of $C_{24}$ + Sc system as compared to the isolated scandium atom in the valence band near the Fermi level. This denotes that some charge has been transferred from Sc-4s orbital to C-2p orbitals of $C_{24}$ fullerene when the scandium atom is decorated to $C_{24}$ fullerene. From the partial density of states picture, it is pretty clear that some charge has been transferred from the scandium's 3d and 4s orbital to the C-2p orbitals of $C_{24}$ fullerene. Due to this charge transfer, the scandium atom is strongly bonded with the $C_{24}$ fullerene molecule.

**Bader charge analysis**

Partial density of states (PDOS) analysis depicts a qualitative picture of charge transfer. We have performed the Bader charge analysis[52] to get the exact amount of charge transferred from the scandium atom to $C_{24}$ fullerene. By performing the Bader charge analysis for $C_{24}$ fullerene and scandium decorated $C_{24}$ fullerene, we have found that 1.44e charge has been transferred from the scandium atom's 3d and 4s orbitals to C-2p orbitals of $C_{24}$ fullerene. This charge transfer is mainly responsible for the strong binding (-3.44 eV) of the scandium atom to the fullerene structure.

**Charge density plot**

We have plotted the charge density difference 2D plot, $\rho\ (C_{24} + Sc) - \rho\ (C_{24})$ as displayed in **Fig. 6(a).** The plot is corresponding to the R-G-B color pattern, and the value of the iso-surface is 0.040e. Green and blue color denote the charge loss and charge gain regions, respectively. The charge density difference plot is consistent with the Bader charge and partial density of states analysis.

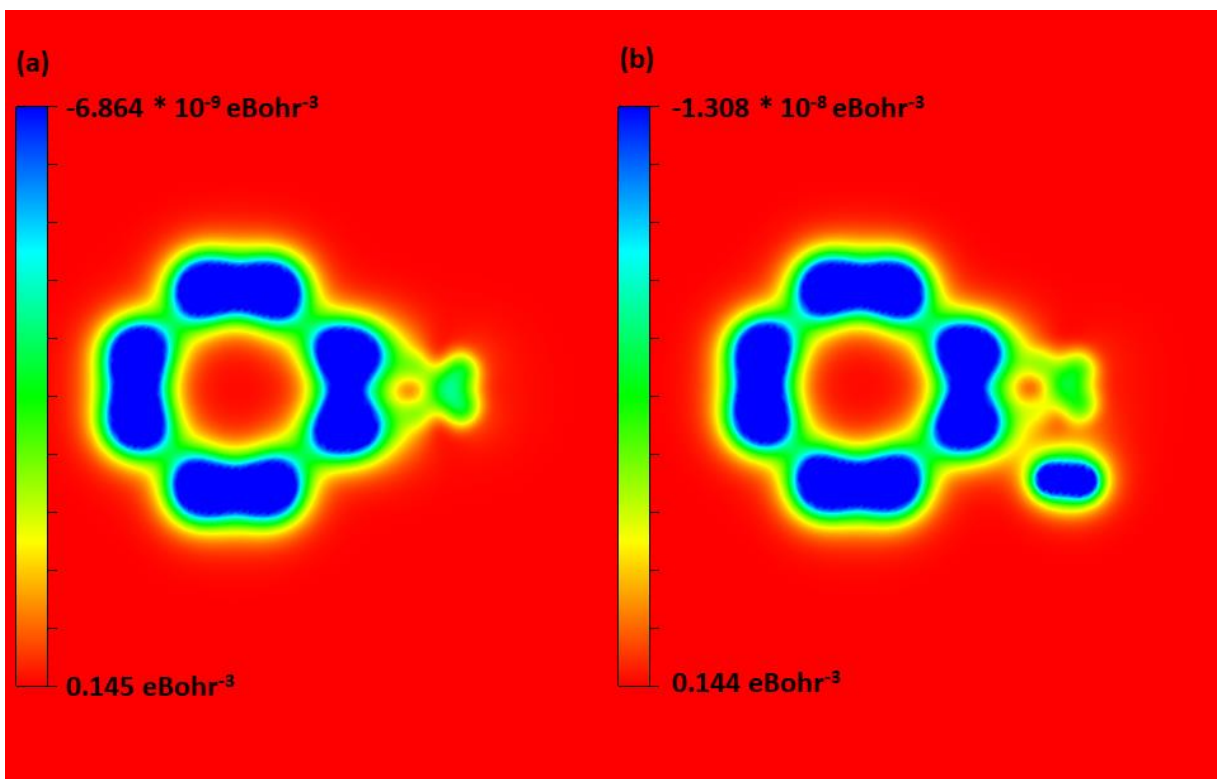

**Fig. 6** Electronic charge density difference 2D plots for (a) $\rho(C_{24} + Sc) - \rho(C24)$ system for isosurface value 0.040e. (b) $\rho(C_{24} + Sc + H_2) - \rho(C_{24} + Sc)$ system for isosurface value 0.036e. Here green color denotes charge loss region and blue color denotes charge gain region.

### 3.3 Adsorption of hydrogen molecules on Sc-decorated $C_{24}$ fullerene

Initially, one hydrogen molecule is kept at almost 2 Å distance above the Sc-decorated $C_{24}$ fullerene structure. We have optimized this structure using GGA exchange-correlation functional and employed the dispersion corrected DFT-D3 method to incorporate the weak van der Waals interactions. The hydrogen molecule is at 2.13 Å distance away from the Sc-decorated $C_{24}$ fullerene after geometry optimization. The H-H bond length increases from 0.75 Å to 0.80 Å after the optimization of $C_{24}$ +Sc+$H_2$ structure. We have presented the optimized structure of $C_{24}$+Sc+$H_2$ composition in **Fig. 7(a).** The binding energy of the first hydrogen molecule on scandium decorated $C_{24}$ fullerene is -0.26 eV which is calculated by using the following formula:

$$(B.E.)_{H_2} = E_{C_{24}+Sc+H_2} - (E_{C_{24}+Sc} + E_{H_2}) \tag{2}$$

Where $E_{C24+Sc+H_2}$, $E_{C24+Sc}$, and $E_{H_2}$ represents the energy of $C_{24}$ + Sc + $H_2$, $C_{24}$ + Sc, and $H_2$ molecules, respectively. The binding energy of the first hydrogen molecule lies in the range of adsorption energy for hydrogen molecules (-0.2 eV to -0.7eV) as guided by DOE-US. We have successively attached hydrogen molecules step by step on $C_{24}$+Sc+$H_2$ structure in a symmetric fashion and observed that scandium decorated $C_{24}$ fullerene can bind up to 6 $H_2$ molecules with suitable binding energy. The average binding energy of the 2$^{nd}$ and 3$^{rd}$ hydrogen molecule is -0.42 eV. After that, 4$^{th}$ and 5$^{th}$ hydrogen molecules are added on $C_{24}$+Sc+3$H_2$ structure in a symmetric pattern, and the average binding energy is found to be -0.37 eV. The binding energy of the 6$^{th}$ hydrogen molecule is -0.28 eV. The binding energies of hydrogen molecules are in the middle of the DOE range (-0.2 to -0.7) eV and most suitable for practical fossil fuel purposes.

The optimized structures of $C_{24}$+Sc+3$H_2$, $C_{24}$+Sc+5$H_2$, and $C_{24}$+Sc+6$H_2$ are displayed in **Fig. 7(b), Fig. 7(c),** and **Fig. 7(d),** respectively.

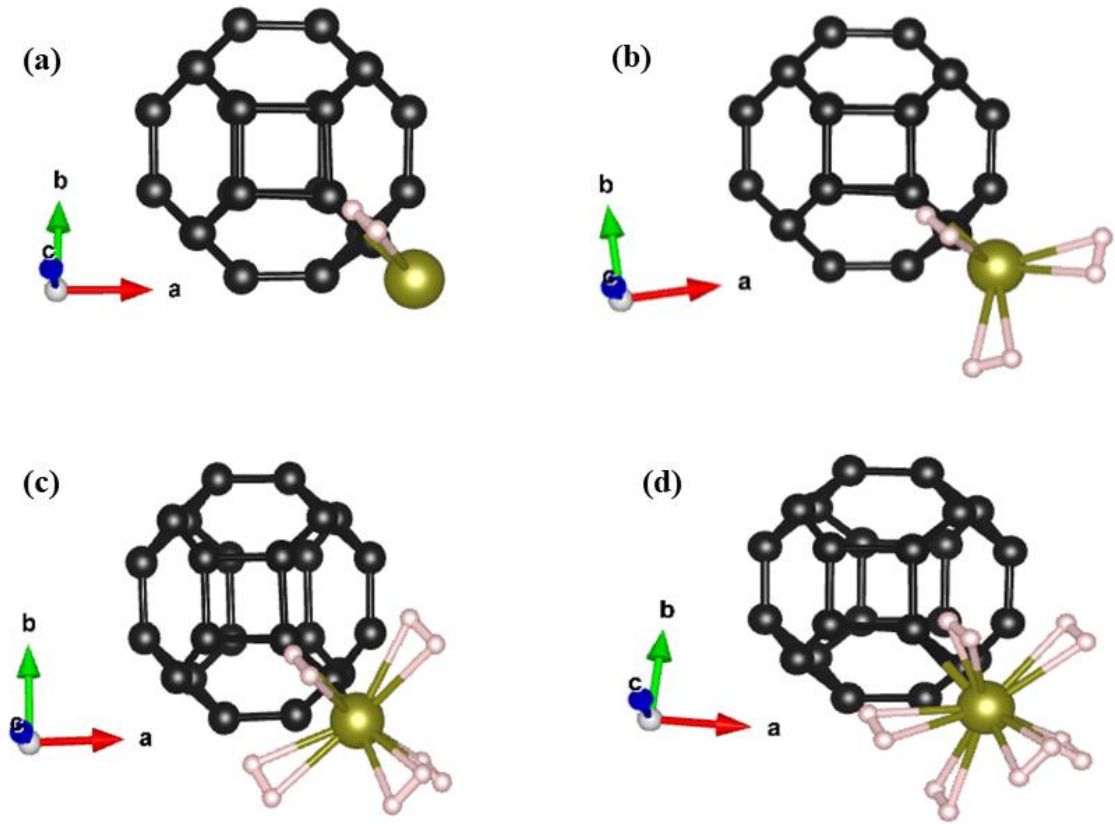

**Fig. 7** Optimized structure of (a) $C_{24}$ + Sc + $H_2$ (b) $C_{24}$ + Sc + $3H_2$ (c) $C_{24}$ + Sc + $5H_2$ (d) $C_{24}$ + Sc + $6H_2$ compositions.

We have presented the average adsorption energy of hydrogen molecules in **Table 1**, employing GGA and GGA+DFT-D3 method. The values of H-H bond length for the adsorbed hydrogen molecules are also presented in **Table 2.**

**Table 1. Average adsorption energy of Sc and $H_2$ molecules with GGA and GGA+DFT-D3 method.**

| Compositions | Adsorption energy (eV) GGA | Adsorption energy (eV) GGA + DFT-D3 |
|---|---|---|
| Fullerene $C_{24}$ | 0 | 0 |
| $C_{24}$ + Sc | -3.44 | - |
| $C_{24}$ + Sc + $H_2$ | -0.26 | -0.30 |
| $C_{24}$ + Sc + $3H_2$ | -0.42 | -0.46 |
| $C_{24}$ + Sc + $5H_2$ | -0.37 | -0.43 |
| $C_{24}$ + Sc + $6H_2$ | -0.28 | -0.39 |
| Average adsorption Energy per $H_2$ | -0.35 | -0.41 |
| Average desorption Temperature | 451 K | 529 K |
| Gravimetric wt % | 13.02 | |

**Table 2. The small change in H-H bond length due to the Kubas-interaction.**

| Compositions | Average bond length (H-H) Å |
|---|---|
| $C_{24}$ + Sc + $H_2$ | 0.7969 |
| $C_{24}$ + Sc + $3H_2$ | 0.8048 |
| $C_{24}$ + Sc + $5H_2$ | 0.7923 |
| $C_{24}$ + Sc + $6H_2$ | 0.7886 |

### 3.4 Bonding mechanism and charge transfer between hydrogen molecules and scandium decorated $C_{24}$ fullerene

In this section, we explain the orbital interactions and charge transfer mechanism between the adsorbed hydrogen molecules and scandium decorated $C_{24}$ fullerene structure. We have also explained the nature of the interactions present between the scandium atom and hydrogen molecules.

**Kubas-type of interaction**

The average adsorption energy of hydrogen molecules attached to the scandium decorated $C_{24}$ fullerene structure is -0.35 eV. The average adsorption energy seems to be more than the physisorption and lesser than the chemisorption process. The hydrogen-hydrogen bond length increases slightly when the hydrogen molecules are attached to the scandium atom of the $C_{24}$ + Sc structure. The hydrogen-hydrogen bond length increases from 0.75 Å to 0.80 Å for the first hydrogen molecule attached with the $C_{24}$ + Sc structure. The change in H-H bond length is very small, and the hydrogen atoms are intact in the molecular form. This slight elongation of H-H bond length is due to the Kubas-type of interaction. In Kubas interaction, some charge transfers from the filled σ (Highest occupied molecular orbital) orbitals of hydrogen molecule to the vacant 3d orbitals of scandium atom take places, and subsequently, there is some back charge transfer from the filled d orbitals of scandium to the vacant σ* (Lowest unoccupied molecular orbital) orbitals of the hydrogen molecule. In this charge donation and back donation process, some small net charge is gained by the 1s orbital of hydrogen, which is responsible for the slight elongation in H-H bond length. Therefore, the bonding of hydrogen molecules with the scandium atom is mainly due to the Kubas interaction along with the weak van der Waals interactions. By performing the relaxation calculations for the Zr and V decorated $C_{24}$ fullerene, we have found that the Zr and V atoms also get adsorbed on $C_{24}$ fullerene with binding energy -3.82 eV and -2.73 eV, respectively. The relaxed structures of Zr and V decorated $C_{24}$ fullerene are presented in **Fig. S1**. We have also calculated the binding energy of the first hydrogen molecule attached on the $C_{24}$ + Zr and $C_{24}$ + V compositions. The relaxed structures of $C_{24}$ + Zr + $H_2$ and $C_{24}$ + V + $H_2$ are displayed in **Fig. S2.** The binding energy of the first hydrogen molecule adsorbed on Zr and V decorated $C_{24}$ fullerene is -0.50 eV and -0.46 eV, respectively lies in the range specified by the Department of Energy, United-States. So transition metals other than Sc for example: Zr, V, can also bind the hydrogen molecules through Kubas interactions, however we have chosen the Sc atom because it is the lightest transition metal element and possesses sufficient number of empty d-orbitals to bind the hydrogen molecules through Kubas interactions[33].

**Partial density of states (PDOS) analysis**

To explain the orbital interactions and charge transfer between hydrogen molecules and the scandium atom of the $C_{24}$ + Sc system, in **Fig. 8(a),** we have presented the PDOS of H-1s orbital for isolated hydrogen molecule while in **Fig. 8(b),** the PDOS of H-1s orbital is plotted for $C_{24}$ + Sc + $H_2$ composition. There is some enhancement in the states in the valence band of $C_{24}$ + Sc + $H_2$ composition, as displayed in **Fig. 8(b).** This enhancement in the states is due to some charge gained by hydrogen 1s orbital when hydrogen molecule is attached to $C_{24}$ + Sc system. We have also plotted the PDOS of scandium 3d orbitals for $C_{24}$ + Sc and $C_{24}$ + Sc + $H_2$ compositions as presented in **Fig. 8(c)** and **Fig. 8(d),** respectively. We can notice some small charge loss in the PDOS of scandium 3d orbitals when hydrogen molecule is attached to scandium. This charge loss is due to the transfer of some charge from the scandium 3d orbitals to the hydrogen 1s orbital when the hydrogen molecule is attached to $C_{24}$ + Sc system. This charge transfer is responsible for the slight elongation in the H-H bond involving Kubas interactions.

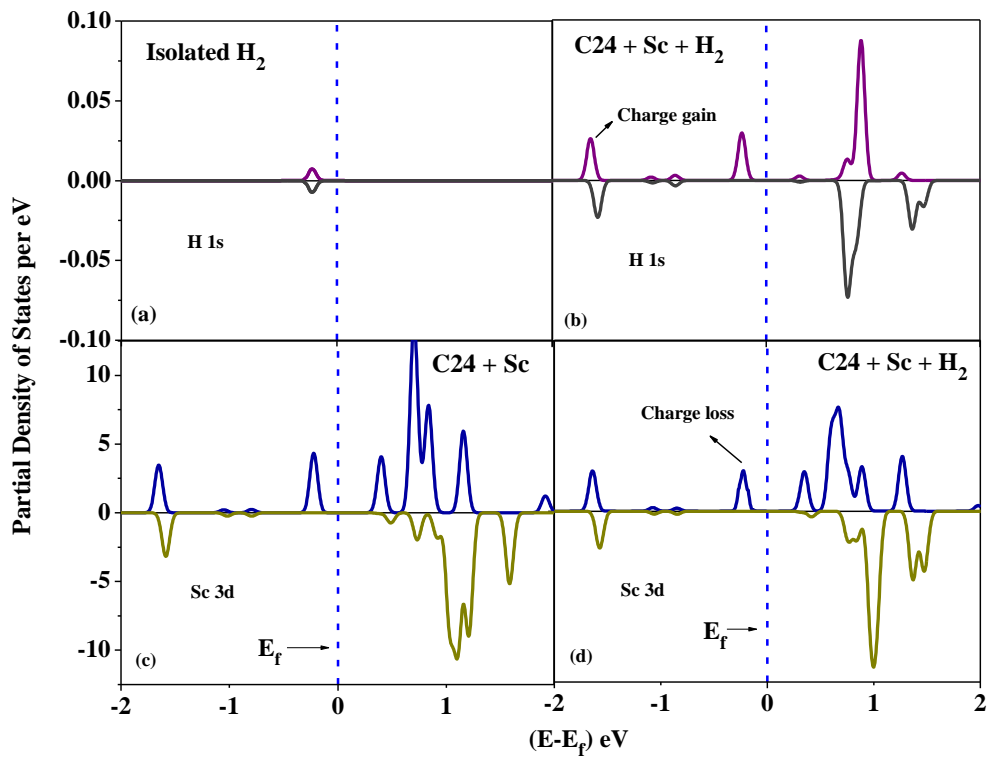

**Fig. 8** Partial density of states for (a) H-1s orbital of isolated $H_2$ molecule. (b) H-1s orbital for $C_{24}$ + Sc + $H_2$. (c) Sc-3d orbitals of $C_{24}$ + Sc. (d) Sc-3d orbitals of $C_{24}$ + Sc + $H_2$. Fermi energy is set at zero energy value.

### Bader charge analysis

To compute the exact amount of charge transfer from the scandium 3d orbitals when hydrogen molecule is attached to the $C_{24}$ + Sc system, we have performed the Bader charge analysis for $C_{24}$ + Sc and $C_{24}$ + Sc + $H_2$ compositions. We have found that a total amount of 1.32e and 0.11e charge transfer from the scandium atom to carbon atoms of $C_{24}$ fullerene and hydrogen atoms of $C_{24}$ + Sc + $H_2$ composition, respectively. This small amount of charge 0.11e, which gets transfer to hydrogen atoms, is responsible for the elongation of the H-H bond.

### Charge density plot

The 2D plot of charge density difference ρ ($C_{24}$ + Sc + $H_2$) − ρ ($C_{24}$ + Sc) is displayed in **Fig. 6(b).** The plot is for the R-G-B color pattern, and the value of iso-surface is 0.036e. We can see that charge has been transferred from the scandium atom to the C-atoms of $C_{24}$ fullerene and H-atoms of the $H_2$ molecule. The charge density difference plot is consistent with the Bader charge and partial density of states analysis.

### 3.5 Calculations of desorption temperature and gravimetric weight percentage (wt %) of hydrogen

Desorption temperature is the temperature at which the adsorbed hydrogen molecules desorb from the system for practical use. The average desorption temperature of hydrogen molecules is calculated using the Van't Hoff formula[65] of desorption temperature:

$$T_d = \left(\frac{E_b}{k_B}\right)\left(\frac{\Delta S}{R} - \ln P\right)^{-1} \qquad (3)$$

In this formula, $k_B$, R, P and, $\Delta S$ are the Boltzmann constant, universal gas constant, atmospheric pressure, and the change in entropy of hydrogen in transition from gaseous to liquid form[66], respectively. $E_b$ is the average adsorption energy of the 6 $H_2$ molecules. The average adsorption energies of hydrogen molecules corresponding to GGA and GGA + DFT-D3 methods are -0.35 eV and -0.41 eV, respectively. The average desorption temperatures of hydrogen are 451 K and 529 K for GGA and GGA + DFT-D3 methods, respectively. The values of the average adsorption energy and desorption temperature are similar to the previously reported values for carbon nanostructures[37,67,68]. The calculated values of average desorption temperature are much higher than room temperature, indicating that the adsorbed hydrogen molecules will not dissociate at room temperature and even in small thermal fluctuations. However, these values of average adsorption energy and desorption temperature are quite suitable for fuel cell applications[69].

In the prediction of gravimetric wt % of hydrogen, it is very important to take care of the possibilities of metal-metal clustering. If the transition metals are very close to each other, they might form a cluster and reduce the hydrogen uptake significantly. In our scandium doped $C_{24}$ fullerene structure, if we put the scandium atom on the top of the tetragon along with the interfacial hexagon position, we can achieve higher hydrogen wt %. To avoid the metal-metal clustering possibilities, we put the scandium atom only in front of the common face of the hexagon, as shown in **Fig 9.**

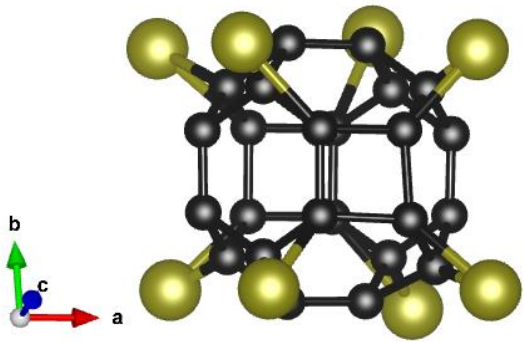

**Fig. 9 Metal loading pattern for $C_{24}$ fullerene. wt % of hydrogen is 13.02.**

By placing the scandium atom in front of the interfacial hexagon position of $C_{24}$ fullerene and considering that one scandium atom will adsorb six hydrogen molecules, we have found 13.02 % of gravimetric hydrogen uptake for our system. The calculated value of hydrogen uptake is much higher than the DOE-US requirement for the hydrogen storage system. The average binding energy of hydrogen, average desorption temperature, and hydrogen uptake for our system are suitable for a practical hydrogen storage system. We have compared these parameters of our system with some of the previous work on similar systems for reference in **Table 3.**

**Table 3. Hydrogen storage parameters comparison for various carbon nanostructures.**

| Metal doped system | Total no. of adsorbed Hydrogen molecules | Average adsorption energy per $H_2$ (eV) | Average desorption temperature (K) | Gravimetric wt % of $H_2$ (%) |
|---|---|---|---|---|
| | | | | |
| Graphene + Ti[18] | 8 | -0.415 | 511.5 | 7.8 |
| Graphyne + Y[39] | 9 | -0.30 | 400 | 10 |
| SWCNT + Ti[34] | 4 | -0.18 | 230 | 8 |
| SWCNT + Y[32] | 6 | -0.41 | 524 | 6.1 |
| $B_{40}$ + Y[40] | 5 | -0.211 | 281 | 5.8 |
| $C_{60}$ + Sc[41] | 4 | -0.30 | - | 7.5 |
| $C_{24}$ + Ti[44] | 4 | -0.33 | 403.5 | 10.5 |
| *$C_{24}$ + Sc* *(Present Work)* | **6** | **-0.35** | **451** | **13.02** |
| **Experimental** | | | | |
| MWCNTs + Pd[45] | - | - | - | 6.0 |
| Graphene + Ni + Al[46] | - | - | - | 5.7 |
| Graphene-Ni Nanocomposites[47] | - | - | - | >6.5 |

### 3.6 Practical feasibility of the system

**Solidity of the structure at desorption temperature**

As we know that the density functional theory calculations are performed at 0 K, we have checked the stability of metal decorated $C_{24}$ fullerene structure at the highest desorption temperature (500K) by performing the *ab-initio* molecular dynamics simulations. The ideal hydrogen storage substrate should adsorb the hydrogen molecules at room temperature, and the desorption temperature of hydrogen should be higher than the room temperature so that the hydrogen molecules remain intact with the structure during thermal fluctuations. It is crucial to check the stability of the hydrogen storage substrate at desorption temperature for the practical use of hydrogen. The *ab-initio* molecular dynamics simulations are carried out in two successive steps. Initially, the scandium decorated $C_{24}$ fullerene structure was kept in a microcanonical (NVE) ensemble for 5ps. In this step, the temperature is raised slowly up to 500 K in the time steps of 1 fs. After that, we have kept the structure in a canonical (NVT) ensemble at a fixed temperature of 500 K for another 3 ps. We have found that the scandium decorated $C_{24}$ fullerene structure is stable, and the change in C-C and C-Sc bond length is negligible after putting the system in the canonical ensemble. The *ab-initio* molecular dynamics snapshot of the structure is displayed in **Fig. 10 (a).** The time evolution of C-Sc bond length (nearest neighbor distance) at 500 K is presented in **Fig. 10 (b).**

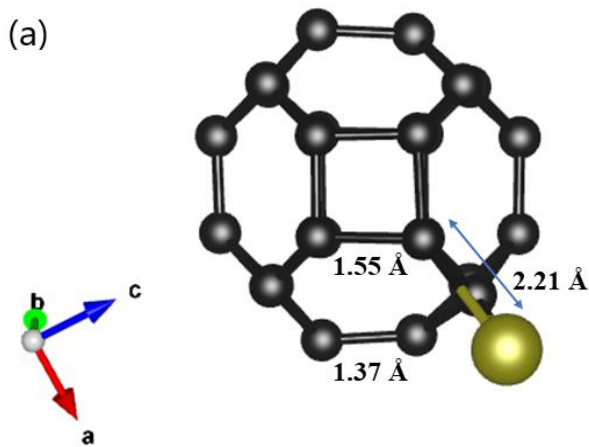

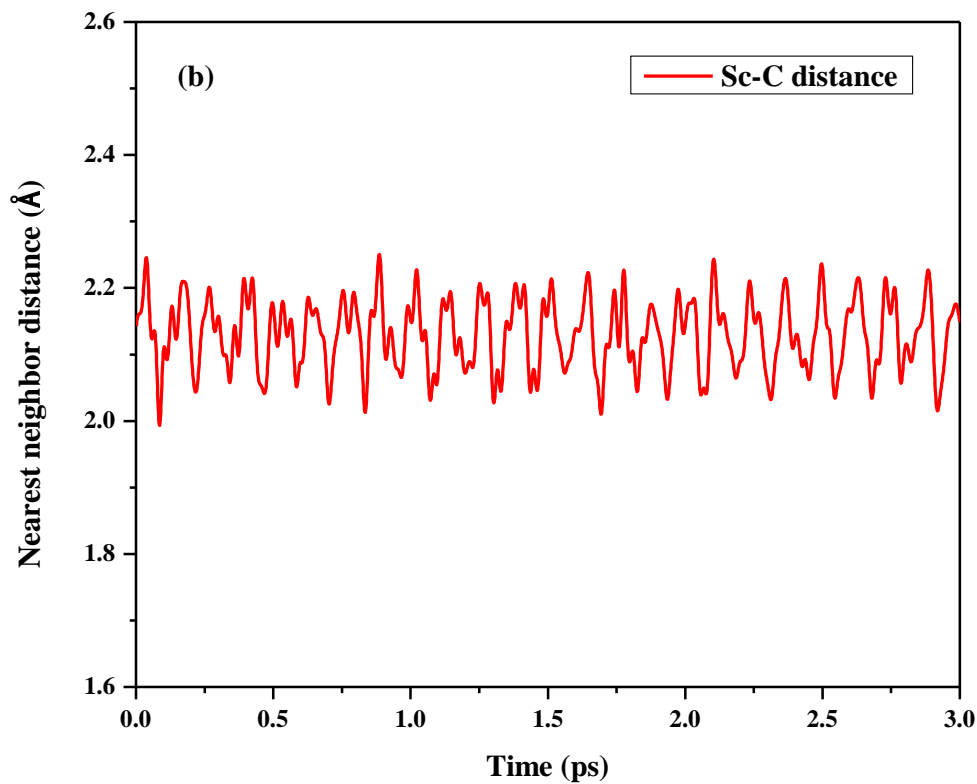

**Fig. 10 (a)** *ab-initio* **Molecular dynamics snapshot of $C_{24}$ + Sc system after putting the system in canonical ensemble at 500 K for 5 ps. (b) Time evolution of the nearest carbon atom distance from scandium atom at 500 K.**

We have observed that the C-Sc bond length is oscillating around the mean value, and the change is very small, which indicates the structural solidity at desorption temperature. Since the scandium atom remains intact with the fullerene structure at the highest desorption temperature, we believe that our system is a practically applicable hydrogen storage system.

**Diffusion energy barrier calculations**

Metal-metal clustering is one of the major issues, which can reduce the hydrogen uptake in a metal decorated carbon nanostructure to a great extent. If the diffusion energy barrier of the transition metal atom is comparable to the thermal energy of the transition metal atom at the highest desorption temperature, then there may be some possibility of metal-metal clustering in the system. Most of the previous studies on hydrogen storage did not include the diffusion energy barrier calculation for the transition metal. This is also one of the important aspects of our work, which assures that the metal-metal clustering may not occur in our system and the system is practically suitable for hydrogen storage.

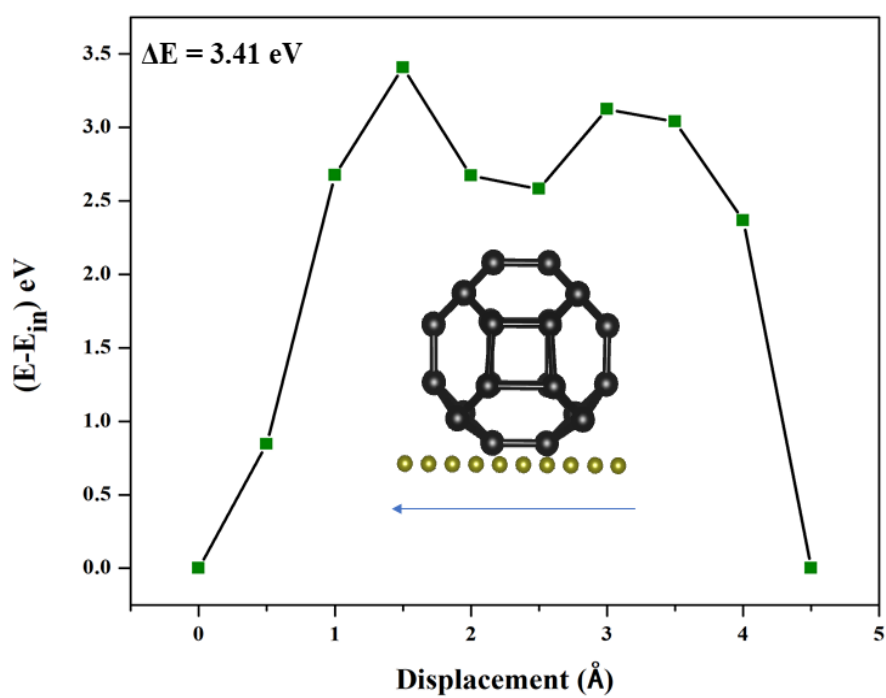

**Fig. 11 Diffusion energy barrier plot for the movement of the Sc-atom. Energy difference of current step energy and initial energy is plotted with respect to the small displacements of Sc atom.**

We calculate the diffusion energy barrier for the scandium atom by displacing the scandium atom slowly from its one stable position. The energy difference between the current step energy and the initial energy of the stable scandium configuration is plotted with respect to the displacement of the scandium atom in **Fig. 11,** and the diffusion energy barrier is calculated. The calculated value of the diffusion energy barrier is much higher (3.41 eV) than the thermal energy of the decorated scandium atom (0.065 eV) at the highest desorption temperature (500 K). The thermal energy of the scandium atom at the highest desorption temperature (500K) is calculated using the following formula:

$$E = \frac{3}{2} k_B T \qquad (4)$$

Where E is the thermal energy of the scandium atom, $k_B$ is the Boltzmann constant, and T is taken to be 500K (more than the desorption temperature). We have found that the thermal energy of scandium atom at 500 K is 0.065 eV, which is much lower than the diffusion energy barrier 3.41 eV of the scandium atom.

## 4 CONCLUSIONS

We have reported the high hydrogen storage capability of scandium decorated $C_{24}$ fullerene using density functional theory calculations. Scandium atom is strongly attached to the $C_{24}$ fullerene (-3.44 eV) due to the charge transfer from the scandium atom to the carbon atoms of $C_{24}$ fullerene. The scandium decorated $C_{24}$ fullerene can adsorb a maximum number of 6 hydrogen molecules with average adsorption energy -0.35 eV and an average desorption temperature of 451 K. The average adsorption energy and desorption temperature of hydrogen are suitable for fuel cell applications, and the hydrogen uptake is 13.02 % for our system. The gravimetric wt % of hydrogen is much higher than the DOE-US requirements. We have performed diffusion energy barrier calculations which indicates that the metal-metal clustering may not take place, ensuring the practical viability of the structure. We have checked the solidity of the structure at the highest desorption temperature by performing the *ab-initio* molecular dynamics simulations. We believe that the scandium decorated $C_{24}$ fullerene is a practically viable high hydrogen storage candidate.

## Acknowledgment

VM would like to acknowledge DST-INSPIRE for providing the fellowship and SpaceTime-2 supercomputing facility at IIT Bombay for the computing time. BC would like to thank Dr. T. Shakuntala and Dr. Nandini Garg for support and encouragement. BC also acknowledge support from Dr. S.M. Yusuf and Dr. A. K Mohanty.

# Supporting Information

# Scandium decorated $C_{24}$ fullerene as high capacity reversible hydrogen storage material: Insights from density functional theory simulations

*Vikram Mahamiya[a], Alok Shukla[a*], BrahmanandaChakraborty[b,c*]*,

[a]Indian Institute of Technology Bombay, Mumbai 400076, India

[b]High pressure and Synchrotron Radiation Physics Division, Bhabha Atomic Research Centre, Bombay, Mumbai, India-40085

[c]Homi Bhabha National Institute, Mumbai, India-400094

email: shukla@phy.iitb.ac.in ; brahma@barc.gov.in

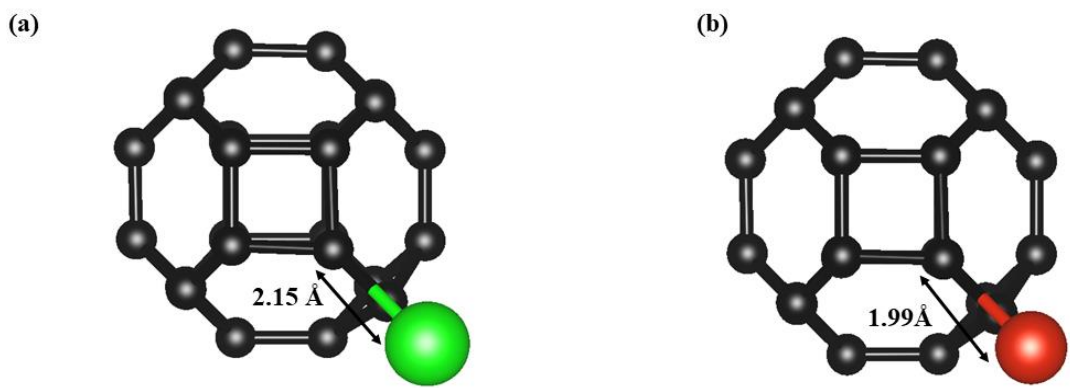

**Fig. S1 Optimized structures of (a) Zr decorated $C_{24}$ fullerene (b) V decorated $C_{24}$ fullerene. Zr and V atoms are placed in front of the common sides of tow hexagons. Black, green and red spheres correspond to C-atom, Zr-atom and V-atom, respectively.**

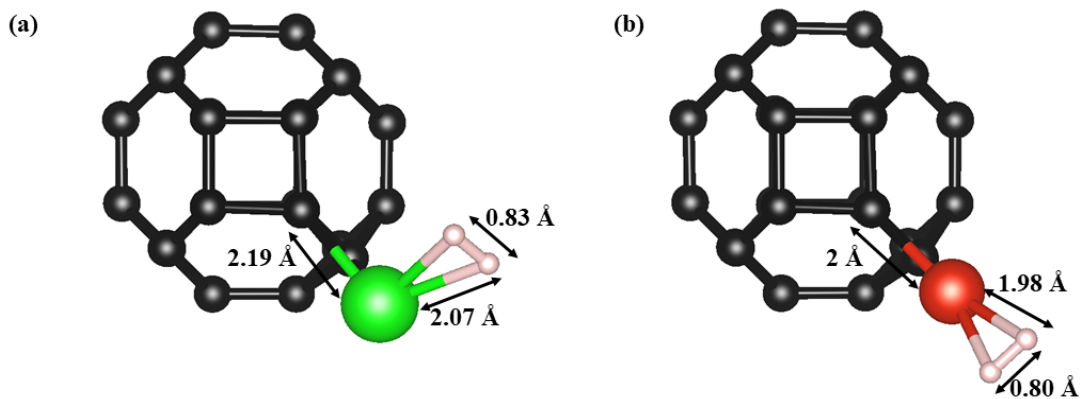

**Fig. S2** Optimized structures of (a) $C_{24}$ + Zr + $H_2$ (b) $C_{24}$ + V + $H_2$. The binding energy of the first hydrogen molecule for Zr and V decorated $C_{24}$ fullerene is -0.50 eV and -0.46 eV, respectively.